\documentclass[11pt]{article}
\pdfoutput=1
\usepackage{fullpage}
\usepackage{xcolor}
\usepackage{cite}
\usepackage{graphicx}
\usepackage{tensor}
\usepackage{amsmath}
\usepackage{amssymb}
\usepackage{subfig}
\usepackage[utf8x]{inputenc} 
\title{}
\date{}
\renewcommand{\vec}[1]{\mbox{\boldmath$ #1 $}}

\def\beq{\begin{equation}}
\def\eeq{\end{equation}}
\allowdisplaybreaks

\begin{document}
\bibliographystyle{utphys}

\newcommand{\hel}{\eta} 
\renewcommand{\d}{\mathrm{d}}
\newcommand{\dd}{\hat{\mathrm{d}}}
\newcommand{\del}{\hat{\delta}}
\newcommand{\ket}[1]{| #1 \rangle}
\newcommand{\bra}[1]{\langle #1 |}

\newcommand{\be}{\begin{equation}}
\newcommand{\ee}{\end{equation}}
\newcommand\n[1]{\textcolor{red}{(#1)}} 
\newcommand{\diff}{\mathop{}\!\mathrm{d}}
\newcommand{\lb}{\left}
\newcommand{\rb}{\right}
\newcommand{\f}{\frac}
\newcommand{\pd}{\partial}
\newcommand{\tr}{\text{tr}}
\newcommand{\fdiff}{\mathcal{D}}
\newcommand{\im}{\text{im}}
\let\caron\v
\renewcommand{\v}{\mathbf}
\newcommand{\T}{\tensor}
\newcommand{\R}{\mathbb{R}}
\newcommand{\C}{\mathbb{C}}
\newcommand{\Z}{\mathbb{Z}}
\newcommand{\msbar}{\ensuremath{\overline{\text{MS}}}}
\newcommand{\DIS}{\ensuremath{\text{DIS}}}
\newcommand{\abar}{\ensuremath{\bar{\alpha}_S}}
\newcommand{\bb}{\ensuremath{\bar{\beta}_0}}
\newcommand{\rc}{\ensuremath{r_{\text{cut}}}}
\newcommand{\Nd}{\ensuremath{N_{\text{d.o.f.}}}}
\newcommand{\red}[1]{{\color{red} #1}}

\newcommand{\Ad}{\dot{A}}
\newcommand{\Bd}{\dot{B}}
\newcommand{\Cd}{\dot{C}}
\newcommand{\Dd}{\dot{D}}
\newcommand{\Ed}{\dot{E}}
\newcommand{\Fd}{\dot{F}}
\newcommand{\depsilon}{\epsilon}
\newcommand{\dsigma}{\bar{\sigma}}

\newcommand{\bphi}{\phi} 
\newcommand{\bB}{B} 
\newcommand{\bH}{H} 
\newcommand{\bsigma}{\sigma} 
\newcommand{\charge}{\tilde{c}} 
\newcommand{\ampA}{\mathcal{A}} 
\newcommand{\ampM}{\mathcal{M}} 

\titlepage

\vspace*{0.5cm}

\begin{center}
{\bf \Large Time-dependent solutions of biadjoint scalar field theories}

\vspace*{1cm} 
\textsc{Kymani Armstrong-Williams\footnote{k.t.k.armstrong-williams@qmul.ac.uk} and
Chris D. White\footnote{christopher.white@qmul.ac.uk}} \\

\vspace*{0.5cm} Centre for Theoretical Physics, School of Physical and
Chemical Sciences, \\ Queen Mary University of London, 327 Mile End
Road, London E1 4NS, UK\\

\end{center}

\vspace*{0.5cm}

\begin{abstract}
Biadjoint scalar field theories appear in the study of scattering
amplitudes and classical solutions in gauge, gravity and related
theories. In this paper, we present new exact solutions of biadjoint
scalar field theory, showing that time-dependent solutions are
possible and analytically tractable. We generalise the theory to
include mass and / or quartic terms, and also a coupling to a constant
background field. This allows for more exact solutions, which make
contact with previous soliton literature. We also find bounded
solutions, in contrast to all known previous examples. Our results may
be useful for the study of non-perturbative aspects of the double copy
between gauge theories and gravity. We also speculate as to their
possible practical applications.
\end{abstract}

\vspace*{0.5cm}

\section{Introduction}
\label{sec:intro}

In recent years, a certain field theory known as {\it biadjoint scalar
  field theory} has appeared in the literature. It is relevant, for
example, for the double copy formalism that relates scattering
amplitudes~\cite{Kawai:1985xq,Bern:2008qj,Bern:2010ue,Bern:2010yg} and
classical
solutions~\cite{Monteiro:2014cda,Luna:2015paa,Ridgway:2015fdl,Bahjat-Abbas:2017htu,Carrillo-Gonzalez:2017iyj,CarrilloGonzalez:2019gof,Bah:2019sda,Alkac:2021seh,Alkac:2022tvc,Luna:2018dpt,Sabharwal:2019ngs,Alawadhi:2020jrv,Godazgar:2020zbv,White:2020sfn,Chacon:2020fmr,Chacon:2021wbr,Chacon:2021hfe,Chacon:2021lox,Dempsey:2022sls,Easson:2022zoh,Chawla:2022ogv,Han:2022mze,Armstrong-Williams:2022apo,Han:2022ubu,Elor:2020nqe,Farnsworth:2021wvs,Anastasiou:2014qba,LopesCardoso:2018xes,Anastasiou:2018rdx,Luna:2020adi,Borsten:2020xbt,Borsten:2020zgj,Goldberger:2017frp,Goldberger:2017vcg,Goldberger:2017ogt,Goldberger:2019xef,Goldberger:2016iau,Prabhu:2020avf,Luna:2016hge,Luna:2017dtq,Cheung:2016prv,Cheung:2021zvb,Cheung:2022vnd,Cheung:2022mix,Chawla:2024mse,Keeler:2024bdt,Chawla:2023bsu,Easson:2020esh,Armstrong-Williams:2024bog,Armstrong-Williams:2023ssz,Farnsworth:2023mff}
in gauge theories with counterparts in gravitational
theories. Potential non-perturbative aspects of this correspondence
have been discussed in
refs.~\cite{Monteiro:2011pc,Borsten:2021hua,Alawadhi:2019urr,Banerjee:2019saj,Huang:2019cja,Berman:2018hwd,Alfonsi:2020lub,Alawadhi:2021uie,White:2016jzc,DeSmet:2017rve,Bahjat-Abbas:2018vgo,Cheung:2022mix,Moynihan:2021rwh,Borsten:2022vtg},
and it is not yet known how generally it holds. Quantities in gauge
theory are related by the so-called {\it zeroth copy} to counterparts
in biadjoint scalar field theory, and this in turn provides crucial
information on what must be left untouched when double-copying gauge
theory objects into gravity ones (see
e.g. refs.~\cite{Borsten:2020bgv,Bern:2019prr,Adamo:2022dcm,Bern:2022wqg,White:2021gvv,White:2024pve}
for pedagogical reviews). This has in turn motivated the study of
exact solutions of biadjoint scalar field theory, which have so far
included spherically symmetric monopole-like
objects~\cite{White:2016jzc,DeSmet:2017rve}, axially symmetric
wires~\cite{Bahjat-Abbas:2018vgo}, and instanton-like solutions in
Euclidean signature~\cite{Armstrong-Williams:2022apo}. Another context
in which biadjoint scalar field theory has arisen is that of geometric
approaches to scattering
amplitudes~\cite{Arkani-Hamed:2017tmz,Arkani-Hamed:2017vfh,Arkani-Hamed:2017mur},
in which the latter are associated with certain polytopes (e.g. the
{\it amplituhedron}) in abstract spaces. This has led to the
consideration of quartic extensions of the original (cubic)
theory~\cite{Banerjee:2018tun,Kalyanapuram:2019nnf,Aneesh:2019cvt,Srivastava:2020dly},
as well as massive variants~\cite{Jagadale:2022rbl} (see
also~\cite{Moynihan:2021rwh}). Gauged versions of biadjoint scalar
theory (including coupling to an external current) have been
considered in ref.~\cite{Cheung:2021zvb}.

Given the relative paucity of results in biadjoint theory compared
with other theories, it is potentially useful to derive new exact
classical solutions, thereby continuing the work of
refs.~\cite{White:2016jzc,DeSmet:2017rve,Bahjat-Abbas:2018vgo,Armstrong-Williams:2022apo}. After
all, the discovery of exact solutions in biadjoint theory mimics the
derivation of soliton solutions of gauge and gravity theories (see
e.g. refs.~\cite{Weinberg:2012pjx,Manton:2004tk,Belinski:2001ph} for
useful compendia), and one might therefore hope that it becomes
possible to match up exact solutions in biadjoint, gauge and gravity
theories, providing a hitherto unrealised non-perturbative realisation
of the classical double copy (see ref.~\cite{Bahjat-Abbas:2020cyb} for
a valiant but unsuccessful attempt). With this in mind, we will
present a number of new solutions in this paper, making contact with
previous soliton literature where possible. Our new solutions are
notable in that they have an explicit time dependence, and represent
travelling excitations akin to plane waves, albeit with a non-trivial
field profile transverse to the wave direction.

The structure of our paper is as follows. In section~\ref{sec:ansatz},
we will present a general biadjoint scalar field equation, which
includes the various extensions to the canonical massless cubic theory
mentioned above. We will present our ansatz for solving this equation,
inspired by e.g. ref.~\cite{Frasca:2009bc}. In
section~\ref{sec:results}, we will present explicit results in closed
form, for increasingly complex cases of the field equation. We discuss
our results and conclude in section~\ref{sec:conclude}.

\section{(Generalised) biadjoint scalar field theory}
\label{sec:ansatz}

Our starting point is to consider the general biadjoint scalar field
Lagrangian
\begin{align}
  {\cal L}&=\frac12 (\partial^\mu \Phi^{aa'})(\partial_\mu\Phi^{aa'})
  -\frac12
  m^2\Phi^{aa'}\Phi^{aa'}-\frac{y}{3}f^{abc}\tilde{f}^{a'b'c'}
  \Phi^{aa'}\Phi^{bb'}\Phi^{cc'}\notag\\
  &-\frac{\lambda}{12}
  \Phi^{aa'}\Phi^{bb'}\Phi^{cc'}\Phi^{dd'} \left( f^{ebc}\tilde{f}^{e'b'c'} f^{eda}\tilde{f}^{e'd'a'} +  f^{edb}\tilde{f}^{e'd'b'} f^{eca}\tilde{f}^{e'c'a'}+  f^{ecd}\tilde{f}^{e'c'd'} f^{eba}\tilde{f}^{e'b'a'} \right)\notag\\
  &+\Phi^{aa'}J^{aa'}.
  \label{Ldef}
\end{align}
Here $\Phi^{aa'}$ are components of a doubly matrix-valued scalar field
\begin{displaymath}
  {\bf \Phi}=\Phi^{aa'}{\bf T}^a\otimes \tilde{{\bf T}}^{a'},
  \label{Phidef}
\end{displaymath}
valued in two Lie algebras with generators $\{{\bf T}^a\}$ and
$\{\tilde{\bf T}^{a'}\}$, and structure constants defined via
\begin{equation}
  [{\bf T}^a,{\bf T}^b]=if^{abc}{\bf T}^c,\quad
    [\tilde{\bf T}^a,\tilde{\bf T}^b]=i\tilde{f}^{abc}\tilde{\bf T}^c.
  \label{fdef}
\end{equation}
Furthermore, $m$ is a mass for the field, and $y$, $\lambda$
constitute independent coupling constants associated with cubic and
quartic interactions respectively. Note that our choice for the sign
of the cubic term differs from
e.g. refs.~\cite{White:2016jzc,DeSmet:2017rve,Bahjat-Abbas:2018vgo,Armstrong-Williams:2022apo},
for later convenience. We have chosen to define the quartic term
solely using the structure constants, which mimics the quartic term of
Yang-Mills theory. Unlike the gauge theory case, however, we have
instigated a separate coupling constant $\lambda$ for generality. We
have written the quartic interaction as a sum over three different
biadjoint colour structures, analogous to how the Feynman rule for the
quartic vertex in Yang-Mills theory is often written as a sum over
three contributions, each of which looks like the colour structure of
two cubic vertices. This plays an important role in the double copy
for scattering amplitudes, but will be irrelevant for our purposes,
such that one may instead consider the Lagrangian
\begin{align}
  {\cal L}&=\frac12 (\partial^\mu \Phi^{aa'})(\partial_\mu\Phi^{aa'})
  -\frac12
  m^2\Phi^{aa'}\Phi^{aa'}-\frac{y}{3}f^{abc}\tilde{f}^{a'b'c'}
  \Phi^{aa'}\Phi^{bb'}\Phi^{cc'}\notag\\
  &-\frac{\lambda}{4}
  f^{ebc}\tilde{f}^{e'b'c'} f^{eda}\tilde{f}^{e'd'a'}
  \Phi^{aa'}\Phi^{bb'}\Phi^{cc'}\Phi^{dd'}  +\Phi^{aa'}J^{aa'},
  \label{Ldef2}
\end{align}
obtainable from eq.~(\ref{Ldef}) by relabelling dummy
indices. Finally, in both eqs.~(\ref{Ldef}, \ref{Ldef2}), we have
included a coupling to an external current $J^{aa'}$, as has been
considered in e.g. ref.~\cite{Cheung:2021zvb}. The above Lagrangians
are expressed in terms of the components $\Phi^{aa'}$ of the biadjoint
field. One may instead use the matrix-valued field of
eq.~(\ref{Phidef}) itself, in which case eq.~(\ref{Ldef2}) corresponds
to
\begin{equation}
  {\cal L}=\frac{1}{T_R\tilde{T}_R}{\rm Tr}\left\{
  \frac12\partial^\mu{\bf \Phi}\partial_\mu{\bf\Phi}-\frac{m^2}{2}{\bf \Phi}^2
  +\frac{y}{3}{\bf \Phi}\left[\left[{\bf \Phi},{\bf \Phi}\right]\right]
  -\frac{\lambda}{4}\left[\left[{\bf \Phi},{\bf \Phi}\right]\right]
  \left[\left[{\bf \Phi},{\bf \Phi}\right]\right]+{\bf \Phi}{\bf J}
  \right\},
  \label{Lmat}
\end{equation}
where we have introduced normalisation constants for the generators in
their given representations:
\begin{equation}
  {\rm Tr}\left[{\bf T}^a {\bf T}^b\right]=T_R\delta^{ab},\quad
  {\rm Tr}\left[\tilde{\bf T}^{a'}
    \tilde{\bf T}^{b'}\right]=\tilde{T}_R\delta^{a'b'}.
  \label{TRdef}
\end{equation}
We have also introduced the double bracket
\begin{equation}
  \left[\left[\Phi,\Phi\right]\right]^{aa'}
  =-f^{abc}\tilde{f}^{a'b'c'}\Phi^{bb'}\Phi^{cc'},
  \label{doublebracket}
\end{equation}
which generalises the conventional Lie bracket in having two
simultaneous structure constants associated with two different Lie
algebras. Note that, to the best of our knowledge, an explicit
Lagrangian for quartic biadjoint scalar theory has not appeared before
in the literature. The closest statement in this regard may be found
in ref.~\cite{Banerjee:2018tun}, which applied amplituhedron-like
methods to quartic biadjoint amplitudes, remarking that the amplitudes
presumably corresponded to a theory possessing a quartic term of the
form we have adopted in eq.~(\ref{Lmat}). 

The field equation arising from eq.~(\ref{Ldef2}) is
\begin{align}
  (\partial^2+m^2)\Phi^{aa'}+yf^{abc}\tilde{f}^{a'b'c'}\Phi^{bb'}\Phi^{cc'}
  +\lambda f^{ebc}\tilde{f}^{a'b'c'}f^{eda}\tilde{f}^{e'd'a'}\Phi^{bb'}
  \Phi^{cc'}\Phi^{dd'}=J^{aa'},
  \label{EOM}
\end{align}
where our choice of numerical constants and signs in eqs.~(\ref{Ldef},
\ref{Ldef2}) has resulted in simpler coefficients in the equation of
motion. Equation~(\ref{EOM}) reduces to the theory studied in
refs.~\cite{White:2016jzc,DeSmet:2017rve,Bahjat-Abbas:2018vgo,Armstrong-Williams:2022apo}
upon choosing $m=\lambda=J^{aa'}=0$ and reversing the sign of $y$. It
is instructive to note, however, a shared property of eq.~(\ref{EOM})
and the simpler case of massless vacuum cubic biadjoint theory: if one
chooses a factorised ansatz of the form
\begin{equation}
  \Phi^{aa'}(x)=\chi^{a}(x)\eta^{a'}(x),
  \label{product}
\end{equation}
the non-linear terms in the equation of motion vanish. Thus,
non-linear solutions can only arise if there is some sort of
cross-talk between the two types of colour charge. With this in mind,
let us choose a common Lie algebra for both sectors of the biadjoint
theory, and make the ansatz
\begin{equation}
  \Phi^{aa'}(x)=\delta^{aa'}f(x),
  \label{ansatz1}
\end{equation}
as has been previously used for static spherically symmetric solutions
of the cubic massless theory in
refs.~\cite{White:2016jzc,DeSmet:2017rve}. Here, however, we will seek
time-dependent solutions, corresponding to non-linear generalisations
of plane waves. Inspired by e.g. ref.~\cite{Frasca:2009bc}, we can set
\begin{equation}
  f(x)\equiv f(\xi),\quad \xi=p\cdot x,
  \label{ansatz2}
\end{equation}
where $p^\mu=(p^0,\vec{p})$ is a constant 4-vector. Solutions of the
form implied by eqs.~(\ref{ansatz1}, \ref{ansatz2}) consisting of a
certain field profile moving in the direction $\vec{p}$ with speed
$p^0/|\vec{p}|$ (in natural units), and where the field is constant on
plane surfaces transverse to the direction of motion. The quantity
$\xi$ then plays the role of a comoving coordinate in terms of which
the field profile is stationary. If we also assume a constant
current density of form
\begin{equation}
  J^{aa'}(x)\equiv J\delta^{aa'},
  \label{Jform}  
\end{equation}
then eq.~(\ref{EOM}) implies
\begin{equation}
  p^2 f''(\xi)+m^2 f(\xi)+yT_A f^2(\xi)+\lambda T_A^2 f^3(\xi)=J,
  \label{EOMf}
\end{equation}
where $T_A$ relates to the normalisation of the generators in the
adjoint representation:
\begin{equation}
  f^{abc}f^{a'bc}=T_A\delta^{aa'}.
  \label{TAdef}
\end{equation}
We now see that the ansatz of eqs.~(\ref{ansatz1}, \ref{ansatz2},
\ref{Jform}) indeed solves the equation of motion, provided that
$f(\xi)$ satisfies the second-order ordinary differential equation of
eq.~(\ref{EOMf}). The latter can be usefully furnished with a physical
interpretation by treating it as motion of a particle in a
one-dimensional potential, with ``mass'' $p^2$, ``position'' $f$ and
``time'' coordinate $\xi$. One may also reduce the order of
eq.~(\ref{EOMf}) by noting that the variable $\xi$ does not appear
explicitly. Then, making the standard substitution
\begin{equation}
  g(\xi)=\frac{df(\xi)}{d\xi},
  \label{gdef}
\end{equation}
eq.~(\ref{EOMf}) assumes the form
\begin{equation}
  p^2 g\frac{dg}{df}+m^2f+yT_A f^2+\lambda T_A^2 f^3=J,
  \label{pdp}
\end{equation}
which can be straightforwardly integrated to give
\begin{equation}
  \frac12 p^2\left(\frac{df}{d\xi}\right)^2+V(f)
  =\varepsilon,\quad V(f)=\frac{m^2 f^2}{2}+\frac{y T_A f^3}{3}
  +\frac{\lambda T_A^2 f^4}{4}-Jf.
  \label{totalE}
\end{equation}
In the physical analogue, this plays the role of the equation for the
total energy of the particle, where the constant parameter
$\varepsilon$ expresses the sum of the kinetic and potential
contributions, and constancy of $\varepsilon$ follows from the lack of
explicit dependence on $\xi$ in the equation of motion for
$f(\xi)$. The possible particle motions $f(\xi)$ will then depend on
the form of the potential, as well as the value of the ``energy''
$\varepsilon$ (which in the field theory context is merely an
integration constant that arises upon solving the field
equation). Various possibilities arise, including unstable solutions
associated with global or local maxima of the potential; non-bounded
solutions where the particle enters from, and returns to, infinity;
and (meta-)stable solutions involving oscillation around a local or
global minimum. Let us now turn to specific cases.

\section{Results}
\label{sec:results}

\subsection{Solutions of the cubic theory}
\label{sec:cubic}

\begin{figure}
\centering
  \subfloat[][]
             {\scalebox{0.4}{\includegraphics{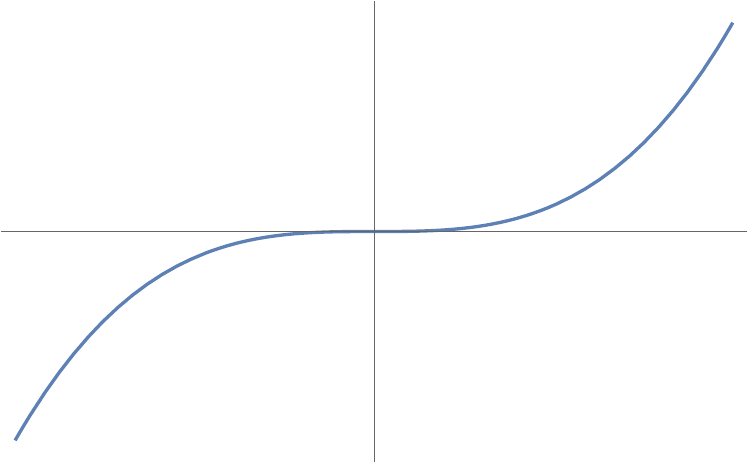}}}
             \qquad\qquad\qquad
    \subfloat[][]
             {\scalebox{0.4}{\includegraphics{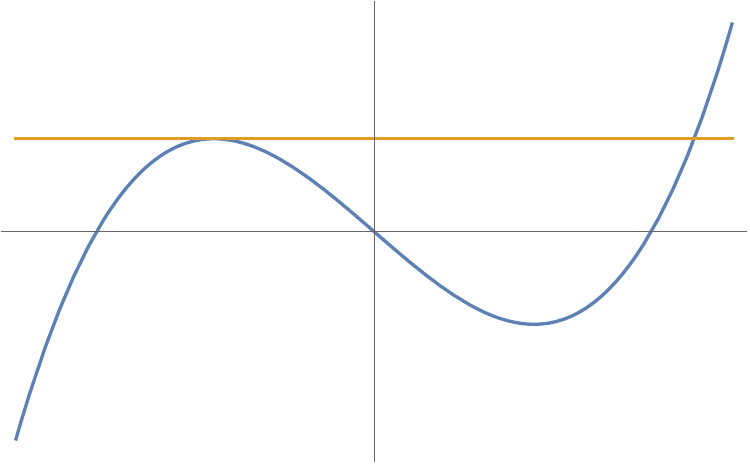}}}
    \caption{(a) Potential energy $V(f)$ corresponding to massless
      cubic biadjoint scalar field theory, for: (a) the vacuum case;
      (b) the non-vacuum case with a constant current density $J>0$.
      In the latter case, oscillatory solutions are possible if the
      energy of the particle is less than the local maximum of the
      potential curve, shown in orange.  }
             \label{fig:cubicpot}
\end{figure}

To make contact with previous literature, we may start by considering
the case $m^2=0$, $\lambda=0$, such that eq.~(\ref{EOM}) reduces to
conventional massless cubic biadjoint scalar field theory. The
potential $V(f)$ is shown for both the vacuum and non-vacuum cases in
figure~\ref{fig:cubicpot}. For the former case, one expects only
non-bounded solutions due to the nature of the potential, which
correspond to the particle climbing the potential hill, reaching a
non-zero height in general, and then rolling back down again. For the
non-vacuum case, a non-zero positive current density leads to a local
minimum, such that one may have bounded oscillatory solutions provided
the ``energy'' $\varepsilon$ is less than the local maximum, as
depicted in figure~\ref{fig:cubicpot}(b).

\begin{figure}
\centering
  \subfloat[][]
             {\scalebox{0.4}{\includegraphics{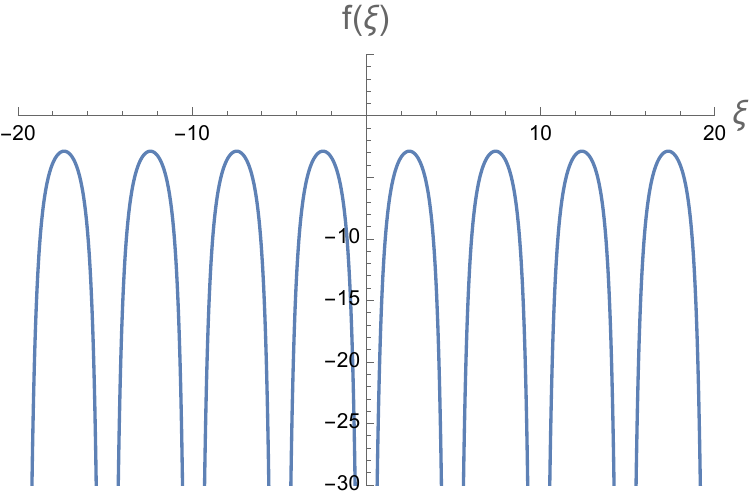}}}
             \qquad\qquad\qquad
    \subfloat[][]
             {\scalebox{0.4}{\includegraphics{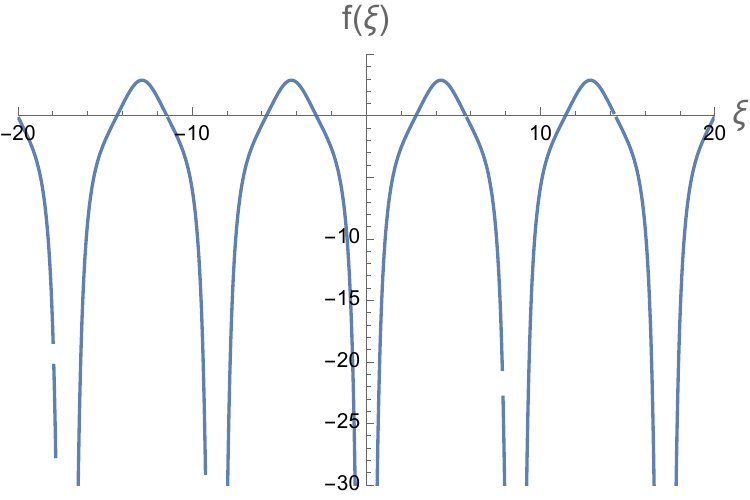}}}
             \qquad\qquad\qquad
    \subfloat[][]
             {\scalebox{0.4}{\includegraphics{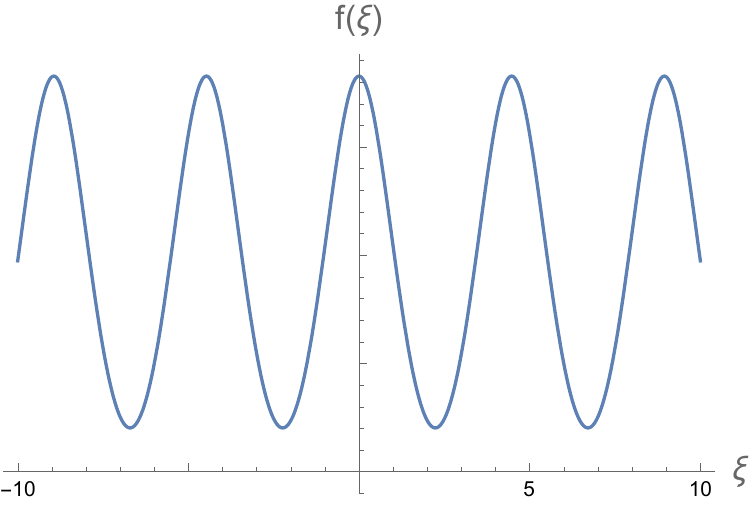}}}
    \caption{Possible solutions for the wave profile $f(\xi)$ in cubic
      massless biadjoint scalar field theory for: (a) $J=0$ and
      $\varepsilon=-4$; (b) $J=0$ and $\varepsilon=4$; (c) $J=2.5$ and
      $\varepsilon=-1$. }
             \label{fig:wpsol}
\end{figure}

That these solutions indeed arise can be seen by comparing
eq.~(\ref{totalE}) in the case $m^2=\lambda=0$ with the defining
differential equation for the Weierstrass elliptic function
$\wp(z,\{g_2,g_3\})$:
\begin{equation}
  \left(\frac{d\wp(z,\{g_2,g_3\})}{dz}\right)^2=4\wp(z,\{g_2,g_3\})
  -g_2\wp(z,\{g_2,g_3\})-g_3.
  \label{wp}
\end{equation}
We then obtain the general solution
\begin{equation}
  f(\xi)=-\frac{6p^2}{yT_A}\wp\left(\xi+c_1,\left\{
  \frac{yT_A J}{3p^4},-\frac{\varepsilon y^2 T_A^2}{18 p^6}
  \right\}\right),
  \label{fsol}
\end{equation}
where $c_1$ is a constant parameter. For $J=0$, the possible forms of
the Weierstrass function are as shown in figure~\ref{fig:wpsol}(a) and
(b), where we have fixed numerical parameters according to $p^2=y=1$,
$\varepsilon=\pm 4$. The form of these solutions can be understood by
appealing to the above classical particle analogy: in the $J=0$ case,
the potential has the form of figure~\ref{fig:cubicpot}(a). We then
expect the particle to roll up the hill, reaching a certain height
before rolling back down again. Depending on the value of the energy,
it may reach different heights before rolling back down.

Returning to the field theory context, the solutions of
figure~\ref{fig:wpsol}(a) and (b) correspond to a generalised plane
wave, where the amplitude diverges periodically along the propagation
direction. The Weierstrass elliptic function behaves like
$(\xi-a)^{-2}$ near a pole $a$, and thus these divergences are
similar to those encountered in the monopole-like solutions of
ref.~\cite{White:2016jzc}. Non-divergent (but metastable) solutions
are possible if the current $J$ is now turned on, given that this can
result in a local minimum in the potential, as shown in
figure~\ref{fig:cubicpot}(b). To find these solutions, one must set
the value of $\varepsilon$ appropriately. One may also use the fact
that, for such solutions, one of the half-periods of the Weierstrass
function in the complex plane of its argument $\xi$ is imaginary
(see e.g. ref.~\cite{Pastras:2017wot} for a pedagogical review). The
real oscillatory solution for $f(\xi)$ is then obtained by shifting
the argument by the imaginary half-period. Figure~\ref{fig:wpsol}(c)
shows an example, for similar parameters as above, but with
$\epsilon=1$ and $J=2.5$.

\begin{figure}
\centering
  \subfloat[][]
             {\scalebox{0.4}{\includegraphics{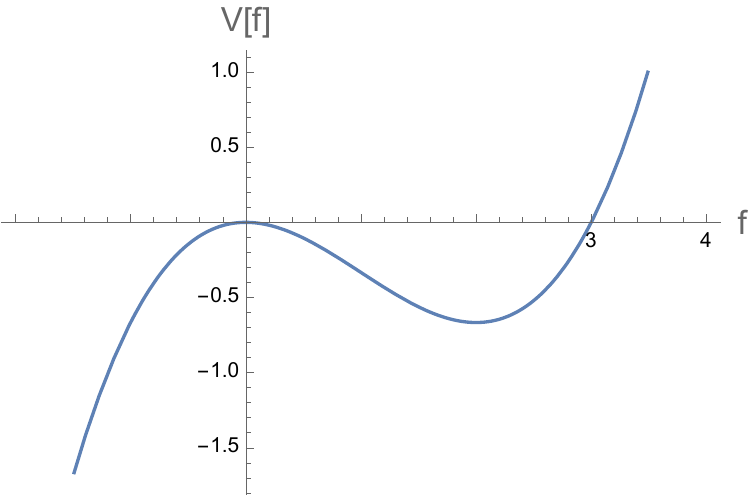}}}
             \qquad\qquad\qquad
    \subfloat[][]
             {\scalebox{0.4}{\includegraphics{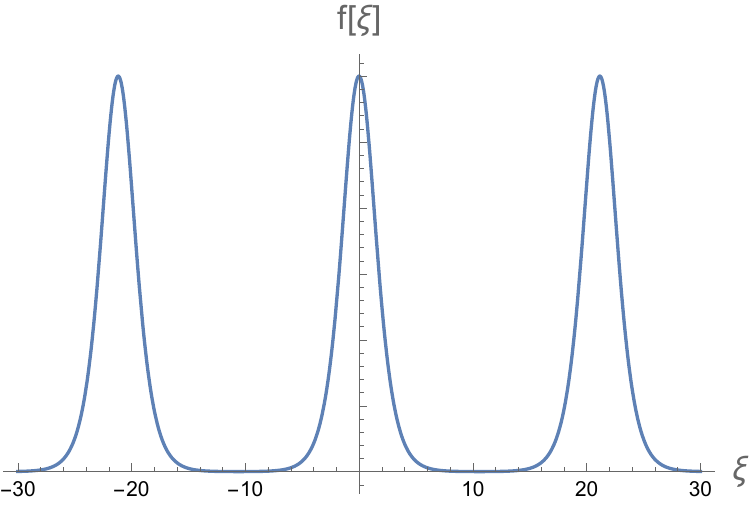}}}
    \caption{(a) Potential energy curve $V(f)$ for massive cubic
      biadjoint scalar theory, with $p^2=y=1$ and $m^2=-1$; (b)
      oscillatory solution obtained numerically, with $f(0)=3$ and
      $f'(0)=0$.}
             \label{fig:cubicmass}
\end{figure}

It may seem somewhat artificial that one can achieve bounded solutions
by turning on a constant source density $J$ filling all space. One can
instead obtain bounded solutions, by using a mass term with
$m^2<0$ to generate the appropriate minimum in the potential, similar
to the case of a Higgs potential. It is then not possible to solve
eq.~(\ref{totalE}) analytically, although numerical solutions can be
straightforwardly obtained, with an example shown in
figure~\ref{fig:cubicmass}. In both the massless and massive cases,
oscillatory solutions play the role of generalised plane waves, where
nonlinear corrections lead to a non-sinusoidal profile along the
longitudinal direction. 

\subsection{Solutions of the quartic theory}
\label{sec:quartic}

\begin{figure}
  \centering
  \subfloat[][]
           {\scalebox{0.4}{\includegraphics{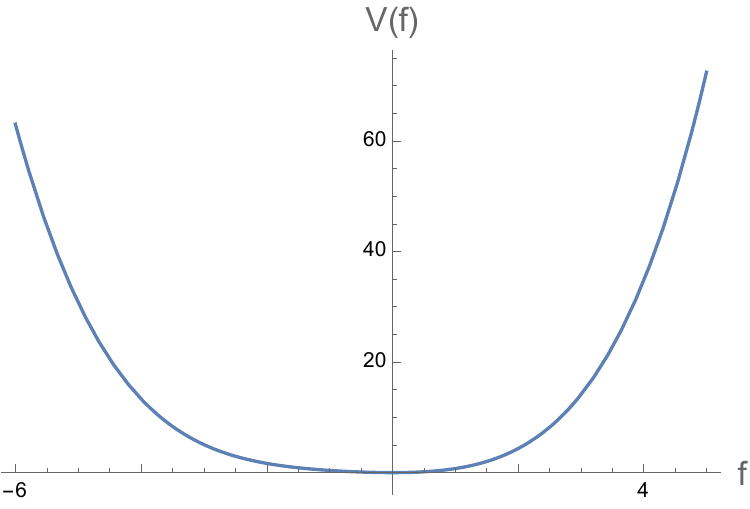}}}
           \qquad\qquad\qquad
           \subfloat[][]
                    {\scalebox{0.4}{\includegraphics{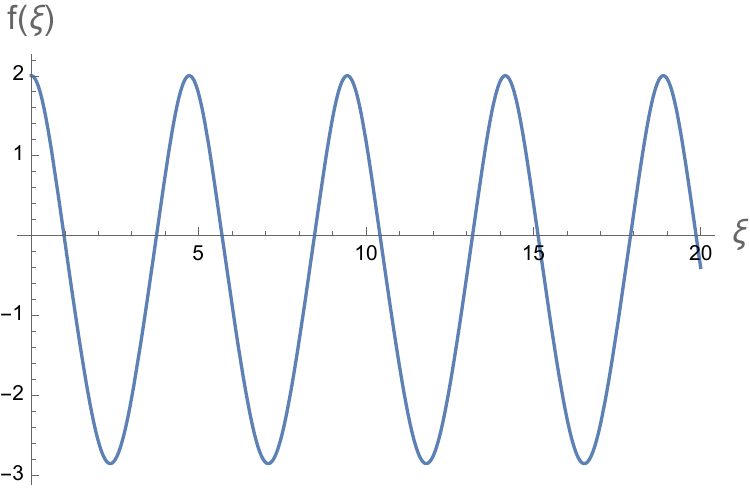}}}
                    \caption{(a) Potential energy curve for massive
                      quartic biadjoint scalar field theory with
                      $p^2=y=m^2=\lambda=1$; (b) oscillatory solution
                      for the wave profile $f(\xi)$ obtained
                      numerically, with $f(0)=2$ and $f'(0)=0$. The
                      solution corresponds, as expected, to
                      oscillations about the minimum of an anharmonic
                      potential well.}
                    \label{fig:numsol}
                    
\end{figure}

Let us now consider the full quartic potential function $V(f)$ of
eq.~(\ref{totalE}), whose general solution $f(\xi)$ can be defined
implicitly via the relation
\begin{equation}
  \xi=\xi_0+\int_{f(\xi_0}^{f(\xi)}\frac{df}{\sqrt{\varepsilon-V(f)}}.
    \label{xidef}
\end{equation}
Given the quartic polynomial form of $V(f)$, the right-hand side
contains an elliptic integral, which may or may not be tractable in
terms of analytic functions, let alone invertible. However, it is
worth drawing attention to the fact that different qualitative
behaviours are possible in the quartic theory than in the cubic
one. There is now the possibility that the potential function $V(f)$
is bounded from below, with either a single or double minimum (for
vanishing cubic term); or two unequal minima (for non-vanishing cubic
term). These possibilities give rise to oscillating solutions around a
minimum or -- in the degenerate vacuum case -- to solutions that
interpolate between the two vacuum field values. In general, solutions
can only be obtained numerically, and an example is shown in
figure~\ref{fig:numsol}.

\begin{figure}
\centering
  {\scalebox{0.4}{\includegraphics{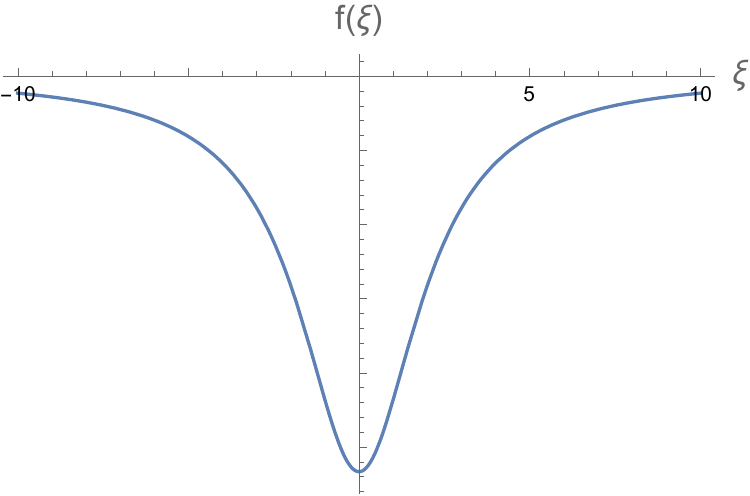}}}
    \caption{Bounded solution for the wave profile $f(\xi)$ in
      massless biadjoint theory (for $y=\lambda=1$), when cubic and
      quartic interactions are both present. }
             \label{fig:bounded}
\end{figure}

Special cases allow
one to make contact with previous literature. For example, choosing
\begin{equation}
  y=0,\quad \varepsilon=-\frac{m^4}{4\lambda T_A^2}
\label{ychoice}
\end{equation}
in eq.~(\ref{totalE}) yields the equation
\begin{equation}
  \left(\frac{d f}{d\xi}\right)^2+\frac{\lambda T_A^2}{2p^2}\left(f^2+
  \frac{m^2}{\lambda T_A^2}\right)^2=0,
  \label{quartic1}
\end{equation}
with general solution
\begin{equation}
  f(\xi)=-\frac{im}{\sqrt{\lambda}T_A}\tanh\left(
  \frac{m(z-c)}{\sqrt{2p^2}}\right)
  \label{fgensol}
\end{equation}
for some constant $c$. A real solution can be obtained upon choosing
imaginary mass $m=i\tilde{m}$ ($\tilde{m}\in{\mathbb R}$), and also
$p^2<0$, in which case eq.~(\ref{fgensol}) reduces (in the static
limit) to the well-known kink solution of quartic scalar field
theory. More generally for vanishing cubic term one finds a solution
in terms of Jacobi elliptic functions:
\begin{equation}
  f(\xi)=\frac{\sqrt{m^4+2c_1\lambda p^2 T_A^2}-m^2}{\lambda T_A^2}
  {\rm sn}\left(\sqrt{\frac{T_A^2\lambda c_1(\xi+c_2)^2}
    {\sqrt{m^4+2p^2T_A^2\lambda c_1}-m^2}},
  \frac{m^2-\sqrt{m^4+2p^2T_A^2\lambda c_1}}
    {m^2+\sqrt{m^4+2p^2T_A^2\lambda c_1}}
      \right),
      \label{frasca}
\end{equation}
which is closely related to the solutions presented in
ref.~\cite{Frasca:2009bc}. Indeed, it is clear that more known
solutions of quartic scalar theories can be naturally embedded within
biadjoint scalar field theories (see e.g. ref.~\cite{Bender:2023cso}
for some recent interesting spherically symmetric solutions). Note
that these particular analytic solutions are only possible if the
cubic term in eq.~(\ref{Ldef2}) vanishes, but the quartic term remains
non-zero. This is made possible in our analysis by the use of
different couplings for the cubic and quartic terms. However, from the
point of view of conventional applications of biadjoint theory
(e.g. in scattering amplitudes related to gauge and gravity theories),
this is rather unnatural. One would expect the cubic term in the
Lagrangian to be present, and the vanishing of this contribution for
non-zero coupling automatically implies the vanishing of the quartic
contribution. Interestingly, one can indeed find analytic solutions in
the massless case with both the cubic and quartic interactions
present. Furthermore, these solutions can be non-oscillatory and
bounded: for $m=J=\varepsilon=0$, eq.~(\ref{totalE}) is solved by
\begin{equation}
  f(\xi)=-\frac{12 p^2 y}{9\lambda p^2 T_A+2 T_A y^2(\xi-c)^2},
  \label{boundedsol}
\end{equation}
shown in figure~\ref{fig:bounded}.

\section{Conclusion}
\label{sec:conclude}

In this paper, we have presented various new analytic solutions of
biadjoint scalar field theories. These theories continue to crop up in
studies of scattering amplitudes and related quantities in quantum
field theory, including being relevant for the relationships between
different types of theory. Exact non-linear solutions may have a role
to play, for example, in probing non-perturbative aspects of the
double copy corespondence between gauge and gravity theories. To this
end, we have generalised the conventional massless cubic biadjoint
theory to include mass and / or quartic terms, as well as coupling to
external currents, all of which variations have appeared in the
literature~\cite{Banerjee:2018tun,Kalyanapuram:2019nnf,Aneesh:2019cvt,Srivastava:2020dly,Jagadale:2022rbl,Moynihan:2021rwh,Cheung:2021zvb}. Our
results extend yet further a catalogue of solutions that has been
gradually growing in the past few
years~\cite{White:2016jzc,DeSmet:2017rve,Bahjat-Abbas:2018vgo,Armstrong-Williams:2022apo}.

Our solutions take the form of nonlinear plane waves, characterised by
a constant four-vector $p^\mu$ that sets the direction of travel and
propagation speed. Specific cases are divergent, mimicing the
divergences seen in static spherically symmetric
solutions~\cite{White:2016jzc}. However, we also find bounded
solutions, either oscillatory or non-oscillatory. The presence of
these qualitatively different behaviours can be understood in all
cases by mapping the differential equation generated by our solution
ansatz to the motion of a point particle in a given potential.

Before concluding, we note that the study of exact solutions in
biadjoint theories is not only of relevance to formal developments in
quantum field theory, but may also have practical applications. In
particular, the addition of a quartic term in the Lagrangian makes the
energy of the theory bounded from below, such that stable
time-dependent solutions can arise. If one restricts to the case of
both Lie groups being SU(2), the biadjoint field generates
infinitesimal rotations of pairs of vectors at the same spacetime
point, and there may well be condensed matter systems for which our
solutions characterise novel collective phenomena. Neither would one
be worried about small-distance divergences in that case, which would
be cut off by the discrete nature of the atomic lattice. We also note
that biadjoint fields involving two different subgroups of the
Standard Model gauge group have been considered in
e.g. ref.~\cite{Carpenter:2024qti}. There may then be interesting
nonperturbative solutions of coupled biadjoint theory, which
generalise the monopoles, instantons and related objects that are
already known. For these reasons, we expect the results of this paper
are not the last word on this matter, and hope to stimulate further
investigation.


\section*{Acknowledgments}

We thank Anja Alfano for communications relating to quartic biadjoint
scalar theory. This work has been supported by the UK Science and
Technology Facilities Council (STFC) Consolidated Grant ST/P000754/1
``String theory, gauge theory and duality''. KAW is supported by a
studentship from the UK Engineering and Physical Sciences Research
Council (EPSRC).

\bibliography{refs}
\end{document}